\documentclass[pra,aps,reprint,superscriptaddress,super,superbib,amsmath]{revtex4-1}
\usepackage{graphicx}
\usepackage[squaren]{SIunits}

\begin{document}

\title{Study of the Quantum Zeno Phenomenon on a Single Solid State Spin}

\author{Janik Wolters} \email[Electronic mail: ]{janik.wolters@physik.hu-berlin.de}
\affiliation{Nano-Optics, Institute of Physics, Humboldt-Universit\"{a}t zu
Berlin, Newtonstr.~15, D-12489  Berlin, Germany}

\author{Max Strau{\ss}}
\affiliation{Nano-Optics, Institute of Physics, Humboldt-Universit\"{a}t zu
Berlin, Newtonstr.~15, D-12489  Berlin, Germany}

\author{Rolf Simon Schoenfeld}
\affiliation{Nano-Optics, Institute of Physics, Humboldt-Universit\"{a}t zu
Berlin, Newtonstr.~15, D-12489  Berlin, Germany}

\author{Oliver Benson}
\affiliation{Nano-Optics, Institute of Physics, Humboldt-Universit\"{a}t zu
Berlin, Newtonstr.~15, D-12489  Berlin, Germany}

\begin{abstract}
The quantum Zeno effect, i.e. the inhibition of coherent quantum dynamics by measurement operations is one of the most intriguing predictions of quantum mechanics. 
Here we experimentally demonstrate the quantum Zeno effect by inhibiting the microwave driven coherent spin dynamics between two ground state spin levels of a single nitrogen vacancy center in diamond.
Our experiments are supported by a detailed analysis of the population dynamics via a semi-classical model.

\end{abstract}

\maketitle
If a quantum system is observed, its coherent dynamics can be significantly slowed down and in case of continuous observation eventually even be frozen~\cite{Misra1977}.
Following a proposal by Cook \cite{Cook1988} this quantum Zeno effect was first demonstrated by the group D. J. Wineland on a microwave induced transitions between ground-state hyperfine levels in ensembles of trapped $^9$Be$^+$ ions by repeated measurements \cite{Itano1990}.
In this letter, we demonstrate the quantum Zeno effect in an individual solid state spin at room temperature. 
We utilize the ground state of a single negatively charged nitrogen vacancy center (NV) in diamond.
The quantum Zeno effect here allows for a detailed study of the intricate interplay of coherent and incoherent dynamics of a single quantum system interacting with a macroscopic environment \cite{DeLange2010}.\\
\begin{figure}[b]
\centering
 \includegraphics[width=0.9\columnwidth]{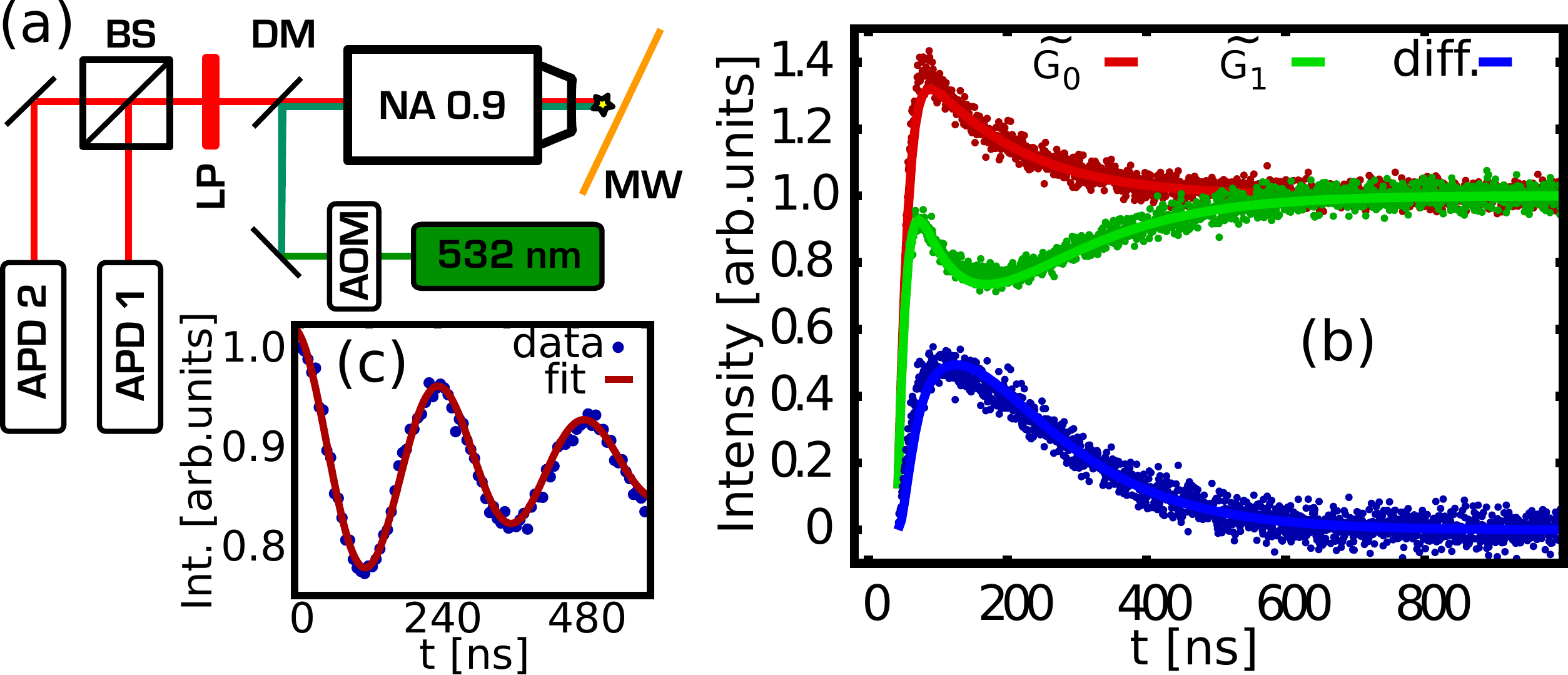}
  \caption{(Color online) (a) Optical setup. 
  An individual NV is excited by a  532~nm cw~laser modulated by an acousto-optic modulator (AOM) through a high NA objective lens. 
  Fluorescence is separated by a dichroic mirror (DM) and additional long pass filter (LP) prior to detection by a Hanbury-Brown and Twiss setup consisting of two avalanche photo diodes (APD) and a beam splitter (BS). 
  Additionally, microwave pulses can be applied via a thin gold wire (MW).
  (b) Fluorescence intensities $I_{\mathrm{G}_{0\,(1)}}(t)$ and the differential signal $I_{\mathrm{G}_{0}}(t)-I_{\mathrm{G}_{1}}(t)$ of the NV   after switch-on of the laser when either the predominantly $\mathrm{m_{s}}=0$ state $\mathrm{\tilde G_{0}}$ or the $\mathrm{m_{s}}=1$ state $\mathrm{\tilde G_{1}}$ was prepared.
    Solid curves are numerical solutions of a rate equation model.
    (c) Measured Rabi oscillations.
  }
  \label{fig:2}
\end{figure}
NVs are frequently used as single photon sources \cite{Schroder2011,Aharonovich2011a}, spin qubits \cite{Ladd2010,Togan2010,Bernien2012}, or bio-compatible sensors \cite{Hall2012}.
In nanodiamonds they can be integrated into various other systems like cells, photonic crystal structures~\cite{Wolters2010,Wolters2012, VanderSar2011} or plasmonic elements~\cite{Schell2011b, Kolesov2009}.
The origin of this versatility is a combination of the NV's level structure, excellent ground state spin coherence and stability of the diamond host lattice~\cite{Jelezko2006,Ladd2010}. 
The electronic level structure is corresponds to the NV's $\mathrm{c_{3v}}$-symmetry \cite{Maze2011,Santori2010a}.
The NV exhibits a triplet ground state $^{3}A_{2}$ with the spin $\mathrm{m_{s}}=0$ sub-level $\mathrm{S_{z}A_{1}}$, and  $\mathrm{m_{s}}=\pm1$ states $\mathrm{S_{x,y}A_{1}}$.
At zero magnetic field these spin levels are split by 2.9~GHz.
The triplet excited state $^{3}E$ has the two-fold degenerate $\mathrm{m_{s}}=0$ sublevels $\mathrm{S_{z}E_{x,y}}$, while the $\mathrm{m_{s}}=\pm 1$ manifold is given by the four-fold degenerate $\mathrm{S_{x,y}E_{x,y}}$.
Using off-resonant excitation with 532~nm levels of same spin are linked by spin-preserving optical transitions near 638~nm.\\
Furthermore, a system of several singlet states with symmetries $^{1}A_{1}$ and  $^{1}E$ exists in between ground and excited triplet states \cite{Maze2011}.
The state $\mathrm{^{1}A_{1}}$ can be reached via inter system crossing (ISC) from the $\mathrm{m_{s}}=\pm 1$ manifold and decays via $\mathrm{E_{1}}$ to the $\mathrm{m_{s}}=0$ ground state.
Via this ISC process the NV  is efficiently polarized within a few excitation-decay cycles. 
In principle the spin state can be measured by exiting the NV once and measuring the NV-spin dependent energy of the subsequently emitted photon.
Thus, a single excitation, decay, and photon emission cycle extracts enough information to unambiguously determine the spin state, i.e. perform a projective measurement. 
In real experiments a low photon collection and detection efficiency \cite{Itano1990} requires repeated measurements. 
In case of NV centers a low Debye-Waller factor and spectral diffusion \cite{Wolters2013} further impede the prediction of the spin state. 
However, since ISC is spin-dependent and the deshelving rate from $\mathrm{E_{1}}$ to $\mathrm{S_{z}A_{1}}$ is low, the $m_{s}=\pm 1$ states appear darker than the $m_{s}=0$ state. 
This correlates the intensity of the fluorescence with the spin state allowing for optical detection of the spin state.\\
Optical spin detection and initialization combined with combined with long coherence times in the mainly spin-free diamond lattice render the NV  ideal to demonstrate coherent spin manipulation.
For example electromagnetic induced transparency \cite{Wrachtrup2006}, simple quantum algorithms~\cite{Shi} and subdiffraction optical magnetometry~\cite{Maurer2010} have been demonstrated.
In Refs.~\cite{Wrachtrup2006, Maurer2010} a 532~nm laser is used to inhibit a population transfer form the  $m_{s}=0$ to $m_{s}=+1$ ground state, but it remains speculative, whether this is due to the quantum Zeno effect, population of the excited state, or simple repumping into the $m_{s}=0$ state.
In this paper we address this question experimentally, as well as by numerical simulations.
Importantly, we demonstrate bidirectional inhibition of the population transfer, i.e. also for the transition from $m_{s}=1$ to $m_{s}=0$.
In this case, simple repumping into the $m_{s}=0$ state \emph{contradicts} the effect, giving clear evidence for the existence of the quantum Zeno effect in this solid state spin system.
As a first step of our investigations we analyze the population dynamics of a selected NV.
In a second step, we use short pulses from a 532~nm laser to initiate a projective measurement and thereby inhibit the coherent microwave driven population transfer between the ground states.
In experiments with ensembles of trapped ions~\cite{Itano1990} a series of measurements was performed as suggested in Ref. \cite{Cook1988}. 
In our case this would lead to significant repumping of the spin state. 
Therefore, we limit ourselves to a single measurement and investigate the influence of the timing of the measurement with respect to the coherent population transfer.
This equivalently allows to study the measurement induced decoherence, i.e. the quantum Zeno effect.\\
\begin{figure}[tb]
\centering
 \includegraphics[width=.37\textwidth]{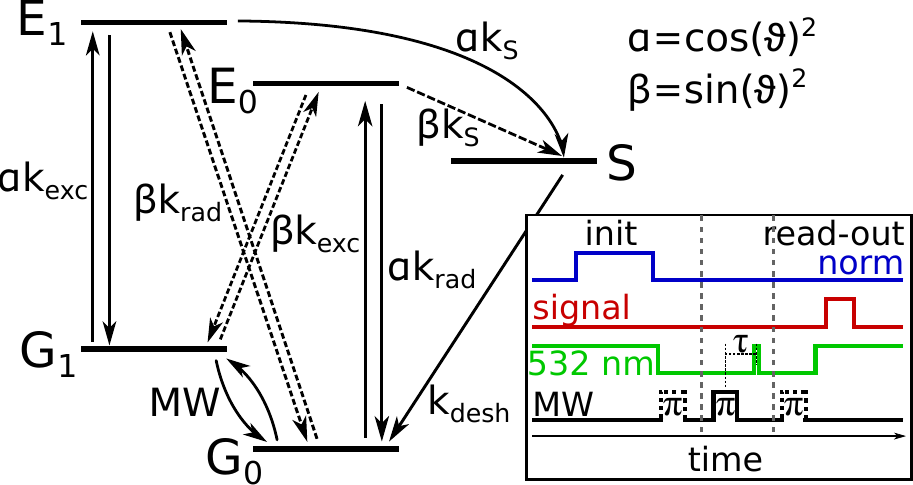}
  \caption{Simplified level diagram of the NV  with strain.
   The $\mathrm{m_{s}}=0\,(1)$ electronic ground state is denoted $\mathrm{G_{0\,(1)}}$, while the corresponding predominantly $\mathrm{m_{s}}=0\,(1)$  excited states are  $\mathrm{E_{0\,(1)}}$.
   The singlet states are merged to the state $\mathrm{S}$. 
   Solid and dashed arrows correspond to allowed and forbidden transitions, respectively. 
  Coherent ground state spin rotations are indicated by MW.
The inset shows the pulse sequence used to measure the Zeno effect. 
   MW denotes the microwave source applying up to 3 $\mathrm{\pi}$-pulses.
    The 532~nm cw laser initializes the NV  at the beginning of the experiment and performs state readout at the end, while a 12~ns short pulse is applied at time delay $\tau$ with respect to the center of the central $\mathrm{\pi}$-pulse. 
   The NV fluorescence intensity is measured within short time windows after and prior to applying the $\mathrm{\pi}$-pulses to obtain the signal and normalization reference.}
  \label{fig:2}
\end{figure}
In our experiment, we use a Type Ib bulk diamond with a solid immersion lens (SIL) produced by focused ion beam milling~\cite{Robledo2011}.
Here strain induces a mixing of NV exited states with different spin orientations.
To account for this, we phenomenologically introduce the spin mixing angle $\vartheta$, resulting in the model depicted in Fig.~2, where we denote the predominantly $\mathrm{m_{s}=0\,(1)}$ excited state $\mathrm{E_{0\,(1)}}$. 
The mixing has two direct implications: Spin non-preserving optical transitions, as well as inter-system crossing from the predominantly $\mathrm{m_{s}=0}$ excited state become possible \cite{Santori2006,Manson2006}. 
According to Fermi's golden rule the rates of the transition allowed in the unstrained NV are reduced by the factor $\alpha=\cos(\vartheta)^{2}$, whereas the formally forbidden transitions have now the rates $k_{x} \beta=k_{x} \sin(\vartheta)^{2}$, with $k_{x}$ being the rate of the corresponding allowed transition. 
To simplify the model, we do not consider the $\mathrm{m_{s}=-1}$ manifold and hence the system can be described by the vector \mbox{$x=\left(\begin{array}{cccccc}G_{0} &G_{1} &\Im(C_{01}) &E_{0} &E_{1} &S\end{array}\right)^{T}$}, where the entries correspond to the level populations, while $\Im(C_{01})$ denotes the imaginary part of the coherence between $G_{0}$ and $G_{1}$.
The dynamics of $x$ is given by $\frac{d}{dt}x=Ax$, with
\begin{eqnarray}
A&=&\left(\begin{array}{cccccc}
-k_{exc} & 0 & -\Omega &  \alpha k_{rad} & \beta k_{rad}&k_{desh} \\
0 & -k_{exc} & \Omega & \beta k_{rad} & \alpha k_{rad} & 0 \\
\frac{i\Omega}{2} & -\frac{i\Omega}{2}  &\Gamma& 0 & 0 & 0 \\
\alpha k_{exc} & \beta k_{exc} & 0 & -k_{rad} & 0 & \beta k_{S} \\
\beta k_{exc} & \alpha k_{exc} & 0 & -k_{rad} & 0 & \alpha k_{S} \\
0 & 0 & 0 & \beta k_{S} & \alpha k_{S} & -k_{desh}\end{array}\right) 
\end{eqnarray}
and $\Gamma= -(1/T_{2}^{*}+k_{exc}) $.

In the investigated diamond, transition rates vary for different centers due to strain and orientation.
To estimate the parameters, a single NV located in the focal point of a SIL was pre-characterized with a homebuilt setup (Fig.~1(a)).
First the ground state spin resonances, which are split up by a small permanent magnetic field by approximately 200~MHz were identified \cite{Wrachtrup2006}. 
Subsequently, coherent Rabi oscillations were driven to measure the oscillation frequency $\Omega=2\pi\cdot4$~MHz, as well as the damping of the oscillation $1/T_{2}^*$ (Fig.~1(c)). 
As the Rabi-frequency is much smaller than the splitting between the $\mathrm{m_{s}=\pm1}$ levels,  individual addressing of the $\mathrm{m_{s}=0}$ to $\mathrm{m_{s}=+1}$ transition is possible.
To estimate the transition rates between different levels, first the $\mathrm{m_{s}=0}$ state was prepared by applying a green cw laser (0.73~mW) for about 5~$\mathrm{\mu s}$. 
About 1~$\mathrm{\mu s}$ after switching off the laser the NV  is assumed to be relaxed into the desired $\mathrm{m_{s}=0}$ state.
Optionally, it can be transferred into the $\mathrm{m_{s}=1}$ state by an additional MW  $\mathrm{\pi}$-pulse.
Subsequently the time dependent fluorescence intensity after switching on the cw laser again was measured (Fig.~1(b)). 
\begin{table}[b]
\caption{Transition rates and parameters deduced by fitting the model to the measurement shown in Fig. 1. 
The error corresponds to one standard deviation confidence interval.\\
}
\begin{center}
\begin{tabular}{cp{0 ex}cp{10 ex}cp{0 ex}c}
\hline\hline
$2\pi/\Omega$&&$(240\,\pm\,7)$\,ns&&$T_{2}^{*}$&&(0.5$\,\pm\,$0.1)\,$\mathrm{\mu s}$\\
$1/k_{exc}$&&($30.5\,\pm\,5$)\,ns&&$1/k_{rad}$&&(13$\,\pm\,$4)\,ns\\
$1/k_{desh}$&&(220$\,\pm\,$60)\,ns&&$1/k_{S}$&&(15.4$\,\pm\,$5)\,ns\\
$\vartheta$&&(12.4$\,\pm\,$3)\degree&&$I_{bg}$&&0.2$\,\pm\,$0.1\\
$\eta_{pol}$&&0.92$\,\pm\,$0.01\\
\hline\hline
\end{tabular}
\end{center}
\label{default}
\end{table}
These dynamics strongly depend on the transition rates, allowing to deduce all remaining free parameters  (Tab.~1) from a fit.
For additional verification, we independently measured the excited states lifetimes with a pulsed laser (PicoQuant), deduced the excitation rate $k_{exc}$ from saturation measurements and the derived contribution of fluorescent background to the signal $I_{bg}$ from autocorrelation measurements. 
From the model a reduced polarization efficiency of the ground state spin  $\eta_{pol}$ can be deduced.
By illumination with green light the bright state $\tilde G_{0}=\eta_{pol} \mathrm{G_{0}}+(1-\eta_{pol}) \exp(i \varphi_{1}) \mathrm{G_{1}}$ with random phase $\varphi_{1}$ is prepared.
A subsequent microwave $\mathrm{\pi}$-pulse transfers the population into the dark state $\mathrm{\tilde G_{1}} = \eta_{pol} \mathrm{G_{1}}+(1-\eta_{pol})  \exp(i \varphi_{2}) \mathrm{G_{0}}$.\\
After determining the NV parameters the quantum Zeno experiment was simulated and experimentally realized. 
The experimental sequence is illustrated in Fig.~2.
First, the NV  is initialized to the bright state $\mathrm{\tilde G_{0}}$ by applying the green laser for about 5~$\mathrm{\mu s}$. 
To initialize the dark $\mathrm{\tilde G_{1}}$ state a subsequent  MW $\pi$-pulse can be applied.
After the initialization the MW pulse is switched on starting a coherent transition from $\mathrm{\tilde G_{0}}$  to $\mathrm{\tilde G_{1}}$ (or from $\mathrm{\tilde G_{1}}$  to $\mathrm{\tilde G_{0}}$).
The pulse is set to a fixed length of 120~ns, i.e. a $\pi$-pulse.
Synchronized to the microwave pulse, a short laser pulse (pulse length 18~ns, peak power 730 $\mu$W) is applied at varying time delay $\tau$. 
With about 1/3 probability this initiates a single cycle of excitation, subsequent spontaneous emission and possible state selective photon detection, i.e. a measurement of the NV center's spin state. 
Even if the final state-selective detection is only done in principle the ground state coherence is destroyed effectively.
Finally, about 300~ns after the microwave pulse the green cw laser is turned on again and the fluorescence of the NV is recorded, giving a measure of the  remaining population of the brighter $\mathrm{\tilde G_{0}}$ state. \\
Independent of the microscopic mechanism, which might be phonon coupling, ISC, or the photon emission and its subsequent absorption,  the laser initiates a measurement of the spin state and destroys the microwave induced coherent polarization, and thereby inhibits the dynamics, similar to experiments with ions~\cite{Itano1990}.
\begin{figure}[tb]
\centering
 \includegraphics[width=.30\textwidth]{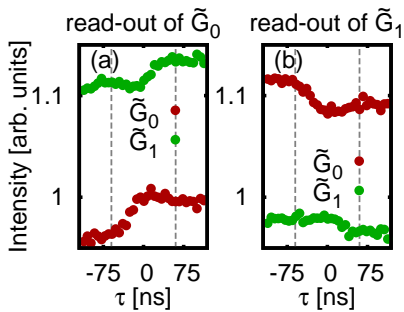}
 \caption{Measurement of the NV fluorescence after the quantum Zeno experiment. 
(a) The NV  is initialized to the bright $\mathrm{\tilde G_{0}}$ (dark $\mathrm{\tilde G_{1}}$) state.
The coherent population transfer during the MW pulse (indicated by dashed lines) is inhibited by a short green laser pulse at time delay $\tau$ with respect to the center of the MW pulse.
Subsequently, the fluorescence as a measure of the occupation of the bright state $\mathrm{\tilde G_{0}}$ is probed.
The Zeno pulse is most effective at $\tau=0$.
(b) Same as (a), but with an additional $\mathrm{\pi}$-pulse before measuring the NV fluorescence, i.e. probing of the dark state $\mathrm{\tilde G_{1}}$.
}
  \label{fig:2}
\end{figure}
In our experiment, this process depends on the time delay between the laser and the MW pulse.
At the center of the MW-pulse the polarization reaches its maximum.
Hence a projective laser pulse at $\tau = 0$ effectively inhibits further coherent dynamics and the final state has a large component of the initial state.
This behavior is clearly visible in Fig.~3(a) as increased (decreased) fluorescence around $\tau = 0$ when the initial state was  $\mathrm{\tilde G_{0}}$  ($\mathrm{\tilde G_{1}}$).
In particular, for an initial  $\mathrm{\tilde G_{1}}$  the decreasing intensity at $\tau=0$ in Fig.~3(a) proves that the effect is not due to repumping of the NV center. 
In this configuration the laser pulse effectively \textit{increases} the population of the $\mathrm{\tilde G_{1}}$ state, resulting in an decreased fluorescence intensity when probing.
Furthermore, due to the large ratio between MW pulse length and excited state lifetime, we can exclude excitation into
the electronic excited state where microwave pulses are off resonant as cause of the effect.\\
\begin{figure}[tb]
\centering
 \includegraphics[width=.28\textwidth]{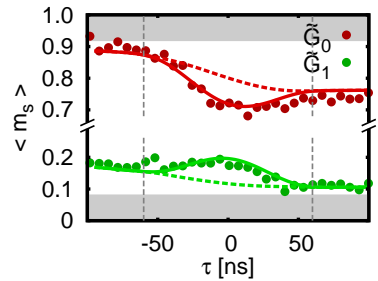}
  \caption{Experimental values (dots) and numerical simulation (solid curves) of the $m_{s}$ spin projection when initially preparing the predominantly  $\mathrm{m_{s}=0}$ (in red) and  $\mathrm{m_{s}=1}$ (in green) level, respectively.
  Time interval between the vertical lines corresponds to the MW pulse, while the shaded areas are inaccessible due to imperfect polarization after initialization. The features centered at $\tau=0$ are whiteness of the Zeno effect. 
Dashed lines indicate the theoretical expectations when decoherence due to the Zeno effect is artificially turned-off.}
  \label{fig:2}
\end{figure}
To get further insight, we repeated the experiment with an additional $\mathrm{\pi}$-pulse before measuring the NV fluorescence (Fig.~3(b)) and calculated  the spin projection  $<\mathrm{m_{s}}>$ from the measured contrast between the intensity with and without this final $\pi$-pulse.
Fig.~4 shows the spin projection together with numerical simulations based on our model.
After preparation of the initial state the polarization efficiency is $\eta_{pol}=0.92$.
Thus, the contrast after preparation of $\mathrm{\tilde G_{0}}$ corresponds to $< \mathrm{m_{s}} >=0.08$, while for the achieved contrast after preparation of $\mathrm{\tilde G_{1}}$  we set $< \mathrm{m_{s}} >=0.92$.
In an ideal Zeno experiment, where the measurement completely destroys the coherence without altering the populations, a detection pulse not overlapping with the MW pulse has no influence. 
A detection within the MW pulse, however, inhibits the coherent dynamics. 
At the extreme case $\tau=0$ the coherent spin transfer is stopped and the final state in an ideal Zeno experiment is a 50/50 mixture of the two spin states. \\
In our experiment the excitation probability is 30\%, well below unity and hence the efficiency of coherent spin transfer is reduced  by about 10\% only. 
Further deviations from the ideal case have several reasons.
There is depolarization via spin non-preserving transitions and repolarization via ISC, as well as the finite lifetime of the excited states.  
Furthermore the limited dephasing time $T_{2}^{*}$ damps the Rabi oscillations, resulting in a small offset towards  $< \mathrm{m_{s}} > = 0.5$, particularly visible at the curve for an initial $\mathrm{\tilde G_{1}}$  for positive $\tau$.
While in the measurement depolarization via spin non-preserving transitions can be neglected, repolarization via ISC is effective when the NV state has a large $\mathrm{G_{1}}$ component at time $\tau$. 
This drives the spin towards $< \mathrm{m_{s}} > = 0$ for positive (negative) $\tau$ in case of initial preparation of $\mathrm{\tilde G_{0}}$ ($\mathrm{\tilde G_{1}}$).
The finite lifetime of the NV excitation has to be considered, when the laser pulse is applied before the end of the microwave pulse, i.e. for  $\tau<60$~ns.
The laser pulse drives the NV to the excited states, where the MW pulse is ineffective. 
However as the MW pulse is long compared to the excited state lifetime and the overall excitation probability is only about 30\%, this effect can be neglected.
To support this analysis we performed a simulation of the experiment, where we artificially introduced an excited state coherence term to turn off the Zeno effect, while all other effects remain unaffected.
While the simulation including the Zeno effect (solid line in Fig.~4)  reproduced the measured data very well, the simulation without Zeno effect (dashed lines in Fig.~4) does not explain the feature centered at $\tau=0$. 
This feature is caused by the destruction of the quantum coherence induced by the green laser pulse initiating the measurement of the spin state, i.e. the quantum Zeno effect.\\
In conclusion we have demonstrated that a short laser pulse initiating a measurement destroys the spin-state coherence of a single NV. 
A MW-driven coherent time evolution can be inhibited in this way.
This can be interpreted as quantum Zeno effect, which was here measured for the first time on a single spin in a solid-state system at room temperature, without the need for any ensemble averaging.
A model including experimentally confirmed parameters was derived to explain the quantum dynamics in detail.
Future work will be devoted to the demonstration of the quantum Zeno effect under spin selective optical excitation, e.g. at cryogenic temperature. 
Here the effect will not only be more pronounced, but is also relevant for recently proposed robust two-qubit quantum gates \cite{Franson2004,Zhou2009,Zhang2011}.
Furthermore, the studied single defect centers are utilized as a scanning quantum emitter probe \cite{Wolters2012,Maletinsky2012,Schell2013}, where modifications of rates (lifetime, decoherence, etc.) by the local environment are monitored as probe signals. Our studies pave the way to exploit full information on the rich coherent and incoherent spin dynamics and thus much deeper insight into interactions of a single quantum system with its mesoscopic environment.

This work was supported by the DFG (FOR 1493).
J.~Wolters acknowledges funding by the state of Berlin (Elsa-Neumann). 
We thank R. Hanson and H. Bernien for the diamond sample and PicoQuant GmbH for collaboration and support.

\bibliographystyle{unsrt}

\end{document}